\documentclass[referee]{raa}            

\usepackage{graphicx,times}             

\begin{document}

   \title{Fast Radio Burst Search: Cross Spectrum vs. Auto Spectrum Method}
   \author{Lei Liu\inst{1,2,3}, Weimin Zheng\inst{1,2,3}, Zhen Yan\inst{1,2}, 
   Juan 
   Zhang\inst{1,3,2}}

  	\institute{$^1$Shanghai Astronomical Observatory, Chinese Academy of 
  	Sciences; liulei@shao.ac.cn\\
	$^2$Key Laboratory of Radio Astronomy, Chinese Academy of Sciences, 
	Nanjing 210008, China\\
	$^3$Shanghai Key Laboratory of Space Navigation and Positioning 
	Techniques, Shanghai 200030, China}

\abstract{The search of fast radio burst (FRB) is a hot topic in current radio 
astronomy study. In this work, we carry 
out single pulse search for a VLBI pulsar observation data set using both auto 
spectrum and cross spectrum search method. The cross spectrum method is first 
proposed in Liu et al. (2018), which maximizes 
the signal power by fully utilizing the fringe phase information of the 
baseline cross spectrum. The auto spectrum search method is 
based on the popular pulsar software package PRESTO, which extracts single 
pulses from the auto spectrum of each station.  
According to our comparison, the cross spectrum method is able to enhance the
signal power and therefore extract single pules from highly RFI contaminated 
data, which makes it possible to carry out FRB search in regular VLBI 
observations with the presence of RFIs.
\keywords{techniques: interferometric --- radio continuum: general --- 
	methods:data analysis --- pulsars: general}
}

   \authorrunning{Lei Liu et al.}
   \titlerunning{Comparison between auto spectrum and cross spectrum based 
   method}
   \maketitle
\section{Introduction}           
\label{sec:intro}
Fast radio burst is a kind of high flux radio burst that is characterized by 
its high dispersion measure (DM) and milliseconds duration. It 
was first reported by Lorimer et al. (2007). Until now, about 20 events are 
discovered with 
large single dish telescopes (Thornton et al. 2013; Spitler et al. 2016) and 
specially designed interferometers (Caleb et al. 2016).
Current studies can almost confirm their extragalactic origin. 
However, their burst mechanism is still not clear. According to Katz (2016), 
the non repeating and repeating burst might have different origins. 

One big challenge in FRB study is their precise localization, which is 
extremely important for discovering their possible afterglow and background 
counterpart in multiple wavelengths. It is expected that various kinds of high 
angular resolution interferometers, e.g., UTMOST (Caleb et al. 2016), CHIME (Ng 
et al. 2017) will be the main stream of FRB search in the 
near future. Besides that, very long 
baseline interferometer (VLBI, Thompson et al. 2001), as the astronomical 
technique that achieves the highest angular resolution, has been used in 
the direct localization of FRB events. The joint observation of VLA, Arecibo, 
EVN and instruments in other wavelengths has revealed the precise 
localization of the repeating burst FRB 121102 and detected the possible 
counterpart in radio and optical bands (Chatterjee et al. 2017; Marcote 
et al. 2017; Tendulkar et al. 2017). Astronomers also try to carry out FRB 
search in legacy VLBI raw data and on going VLBI observations, e.g, the V-FASTR 
project in VLBA (Wayth et al. 2011; Thompson et al. 2011) and the LOCATe 
project in EVN (Paragi 2016).

In general, there are three kinds of VLBI observation data: astrophysics, 
geodetic and deep space exploration. Most of them, if not all, can be used for 
FRB search. Because of the expensive storage, most of these raw data will be 
deleted immediately after correlation. For us, these data are precious and 
deserve further investigation. Our plan is to develop a pipeline to 
carry out FRB search before data deletion. Initially, we chose the popular 
auto spectrum based single pulse search algorithm provided by PRESTO (Ransom 
2001). However, 
soon we realized that the auto spectrum method did not work with the presence 
of RFI. To fully exploit such kind of data, we have to develop new 
method. In Liu et al. (2018), we present a cross spectrum based single pulse 
search method. It utilizes the fringe phase information of baseline cross 
spectrum, so as to maximum the power of single pulse signals. We will introduce 
the method in Sec.~\ref{sec:cross}.

To evaluate the performance of both auto spectrum and cross spectrum based 
single pulse detection methods, we have carried out single pulse search on a 
VLBI 
pulsar observation data set using both methods. The advantage of using pulsar 
data is the arrival time (pulsar phase) of pulsar signal is well known，which 
makes it possible to differentiate if a single pulse is pulsar signal or 
not. 

This paper is organized as follows: In Sec.~\ref{sec:method}, we introduce the 
auto and cross spectrum based single pulse search methods. In 
Sec.~\ref{sec:result}, we present the single pulse detection result using both 
methods. In Sec.~\ref{sec:conclusion}, we summarize the whole work.

\section{The cross spectrum and auto spectrum based methods}\label{sec:method}
\subsection{Cross spectrum method}\label{sec:cross}
The cross spectrum based single pulse search method is first proposed in Liu et 
al. (2018). It takes the idea of fringe fitting in geodetic VLBI data 
postprocessing, which fully utilizes the fringe phase information to maximize 
the signal power (Tahahashi et al. 2000; Cappallo 2014). We make special 
optimizations for the original fringe fitting scheme, so as to achieve higher 
performance and signal power with cross spectrum of millisecond duration. 
The method itself is fully described in Liu et al. (2018). Below we give a 
brief summary:
\begin{itemize}
	\item[a] VLBI correlation of raw data. It is recommended that the 
	station clocks are well adjusted, so that the residual delay is limited 
	to one sample period and the fringe rate is within $10^{-2}$~Hz. The 
	accumulation period of output cross spectrum should be sufficiently small, 
	e.g., 1 millisecond, so as to resolve a typical FRB. 
	\item[b] Dedispersion and construction of time segments. In the cross 
	spectrum method, we carry out incoherent dedispersion on the cross spectrum 
	with millisecond duration. Then several such kind of dedispersed cross 
	spectrum are combined to construct time segment of different window sizes 
	(accumulation period). After this step, several lists of time segments 
	with different window sizes are constructed.
	\item[c] Fringe fitting. For each time segment, we find out 
	the specific multi band delay (MBD) and single band delay (SBD) that 
	maximize the delay resolution function. In the actual implementation, we 
	use a 2D FFT to speedup the search process. 
	\item[d] Single pulse extraction on one baseline. For each time segment 
	list of different window sizes on one baseline, the after fringe fitting 
	signal powers are normalized according to power fluctuation; then single 
	pulses are extracted according to a given threshold. After that these 
	single pulses are filtered in multiple windows to further exclude RFIs. In 
	current scheme, single pulses that are detected on at least 3 windows are 
	selected as candidate signals.
	\item[e] Cross matching candidate signals from multiple baselines. 
\end{itemize}

\subsection{Auto spectrum method}
The famous pulsar search software package PRESTO provides support for auto 
spectrum based single pulse search. The whole process can be divided into 
several steps: 
\begin{itemize}
	\item[a] For each station, carry out incoherent dedispersion on the input 
	auto spectrum. 
	\item[b] Subdivide the auto spectrum into small pieces with given time 
	duration. For each piece, remove the trend and 
	normalize the spectrum with standard deviation; smooth the sample 
	points with multiple down factors.
	\item[c] For sample points in each down factor, pick up candidate signals 
	according to the given threshold. 
	\item[d] Walk through the candidate lists of different down factors, remove 
	candidates that are close to other candidates but are less significant.
	\item[e] Cross match candidate signals detected from multiple stations. Two 
	candidate signals are assumed to match if their time range overlaps with 
	each other.
\end{itemize}
The algorithm of auto spectrum based search method is simple and easy to 
implement. Therefore it is widely used in various kinds of FRB search 
projects. By cross matching candidate signals from 
multiple stations, a significant amount of RFIs can be excluded. However, 
one big disadvantage of this method is, when the radio interference is strong 
or the sensitivity of the station is low, no valid single pulse signals can be 
extracted from the corresponding station. This is clearly demonstrated in 
Sec.~\ref{sec:result}.

One thing we do not mention is the DM search scheme. For both methods, we 
have to divide the target DM search range into several DM bins, and carry out 
single pulse search in each of these DM bins. For auto spectrum method, there 
is an optimized DM search scheme provided by PRESTO. For cross spectrum method, 
the bin width is determined by both the window size and the frequency range as 
proposed in Liu et al. (2018). This work does not involve the DM search. The 
main reason is the DM of the pulsar data set is just 
(26.833~$\mathrm{pc}~\mathrm{cm}^{-3}$), which is too low to carry out 
effective DM search. 

\section{Comparisons of detection result}\label{sec:result}
\subsection{Pulsar data set}
\begin{figure}
\centering
\includegraphics[width=\textwidth]{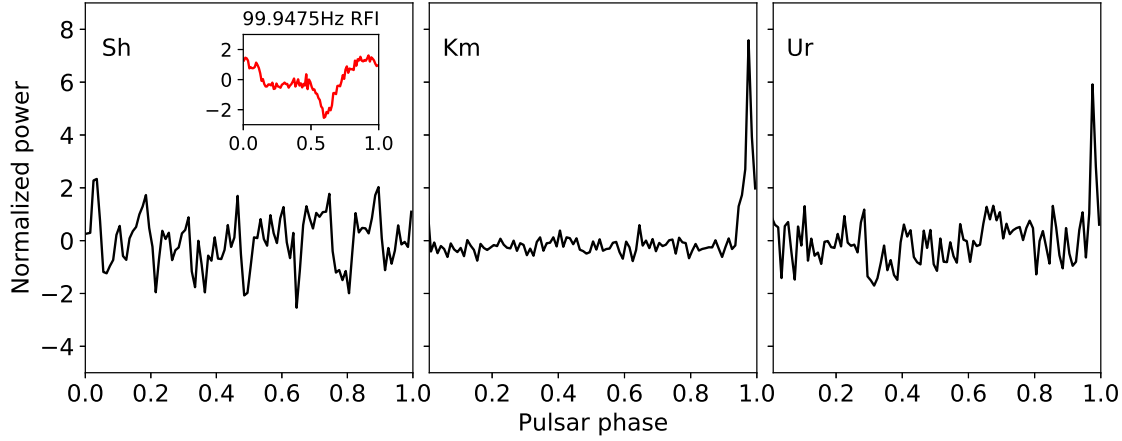}
\caption{Pulse profile of PSR J0332+5434. The profiles are derived by folding 
the data between 10~s and 170~s in scan 73 of CVN observation psrf02. The red 
curve in Sh panel corresponds to the 99.9475 Hz RFI in Sh station. According to 
Liu et al. (2018), the peaks in Km and Ur profile correspond to a pulsar phase 
range from 0.973 to 0.983. \label{fig:profile}}
\end{figure}

The VLBI pulsar data set used in this work is taken from the CVN (Chinese VLBI 
Network, Zheng et al. 2015) pulsar observation of PSR J0332+5434 (Chen et 
al. 2015). The three CVN telescopes, Sh, Km, Ur took part in 
the observation. The SEFD of the three telescopes are 800 Jy, 350 Jy and 560 
Jy, respectively. The target source PSR J0332+5434 is one of the brightest 
pulsar ever known. The average flux is around 0.1 Jy at S band (Kramer et al. 
2013). According to ATNF Pulsar Catalogue (Manchester et al. 2005), the DM 
value is 26.833 pc~cm$^{-3}$ and 
the period is 0.714~s. The 96 MHz observation bandwidth in S band (2192 MHz - 
2288 MHz) is equally divided into six 16 MHz frequency channels. For 
correlation, we use 64 points FFT, which corresponds to 32 frequency points in 
each frequency channel. The observation was carried out on Feb. 15, 2015 and 
lasted for 12 hours. In this 
work, we use pulsar observation scan 69, 71 and 73 for single pulse search. 
Since the starting and ending time of the raw data is 
different for each station and scan, to keep consistency, for each scan, we use 
the data between 10~s and 170~s.

The three panels in Fig.~\ref{fig:profile} demonstrate the pulsar folding 
profiles of PSR J0332+5434 in the three stations. To obtain the profile, we 
first 
carry out time shift on the raw data, so that data from the three stations are 
in 
the same geocentric reference frame. Then those data are Fourier transformed to 
the frequency domain. We calculate the pulsar phase for each frequency point 
and assign it to the corresponding pulsar phase bin. Usually the profile 
appears after enough time of accumulation. The Km and Ur panels show a clear 
pulse profile. As a contrast, the strong 99.9475 Hz RFI makes it impossible to 
extract any valid pulsar signal from Sh station. The peak in Km station
is higher, which corresponds to its higher sensitivity (low SEFD). The 
pulsar phase ranges for the two stations are almost overlapped with each other. 
According to Liu et al. (2018), we set it to 0.973 - 0.983. A single pulse 
is assumed to be a 
``high possibility pulsar signal'' if its time range is overlapped with the 
pulsar phase range. We have to point out that the pulsar phase information 
itself cannot exclude the possibility of false detection. However, it is still 
a good 
criteria to distinguish pulsar signals since single pulses outside this phase 
range are definitely RFIs.

\subsection{Detection results}

In this section, we present the single pulse detection results using both cross 
spectrum and auto spectrum method. 

For the cross spectrum method, we use the CVN software correlator (Zheng et al. 
2010) for VLBI 
correlation. The output accumulation period (AP) is set to 1.024~ms. For fringe 
fitting, we choose 3C273 in scan 293 as the calibration source. For multiple 
windows filtering as described in step d in Sec.~\ref{sec:cross}, we choose the 
window lengths of 4, 8, 16, 24 and 32 APs. 

The single pulse detection result is 
presented in Tab.~\ref{tab:result_cross}. We define the detection accuracy as 
the fraction of high possibility pulsar signals among all the detected signals. 
From the table, Km-Ur baseline yields the highest detection accuracies and the
largest number of high possibility pulsar signals, which is consistent with its 
high sensitivity. In contrast, the detection accuracy of Sh related baselines 
is much lower, which is due to the strong RFI surrounding Shanghai station. In 
Liu et al. (2018), we also present the multiple baselines cross matching 
result. Single 
pulses detected simultaneously on two or three baselines can almost exclude the 
possibility of false detection.

\begin{table}
\centering
\caption{Cross spectrum search results. The number in the parentheses 
corresponds to single pulses of which the pulse time range is overlapped with 
the pulsar phase range (high possibility pulsar signal). 
\label{tab:result_cross}}
	\begin{tabular}{c|ccc}
		\hline\hline
		Scan No.& \multicolumn{3}{c}{Cross spectrum} 		\\
		& Sh-Km & Sh-Ur & Km-Ur \\
		\hline
	69	&	37 (12)	&	33 (2)	&	49 (40)	\\
	71	&	26 (8)	&	35 (3)	&	57 (41)	\\
	73	&	29 (7)	&	34 (4)	&	51 (36)	\\
		\hline\hline
	\end{tabular}
\end{table}

For the auto spectrum method, we first convert the 
Mark5b (Whitney 2003) format raw VLBI observation data to the filterbank format 
which is readable by PRESTO. Raw data are time shifted according to delay 
models, such 
that the filterbank data and the VLBI cross spectrum output are 
in the same geocentric reference frame. Filterbank files are generated for scan 
69, 71 and 73 of Sh, Km, Ur station. Parameters of these filterbank 
files are listed in Tab.~\ref{tab:fb}.

\begin{table}
\centering
\caption{Parameter setting of filterbank files. Low and high channel frequency 
correspond to the frequency in the middle of the respective channel. In the 
filterbank format, file time must be divisible by subint time, therefore it is 
slightly shorter than 160~s. \label{tab:fb}}
\begin{tabular}{ll}
	\hline\hline
	Parameter & Setting \\
	\hline
	Sample time		&	64~$\mu$s \\
	Low channel  	&	2192.5~MHz \\
	High channel 	&	2287.5~MHz \\
	Channel width	&	1~MHz \\
	Channel number	&	96 \\
	Spectra per subint	&	2400 \\
	Spectra	per file&	2498400 \\
	Time per subint	&	0.1536~s\\
	Time per file	&	159.8976~s \\
	Sample bits		&	8 \\
	\hline\hline
\end{tabular}
\end{table}

For single pulse detection with PRESTO, we set a detection threshold of 3 ( 
defaults to 5 in the original program) and a maximum downfactor of 
490\footnote{In the program, the maximum supported value is 300. We modify it 
to 490 to yield a maximum width of 31.36 ms, such that it is comparable with 
the 
maximum window length of 32.768 ms for cross spectrum search in Sec. 
\ref{sec:cross}}. Because of the strong RFI in Sh station, it is 
impossible to detect any meaningful signal. Therefore, we present 
the single pulse detection results for Km and Ur stations only. 

\begin{table}
\centering
\caption{Auto spectrum search results. For comparison, the single pulse 
detection result of Km-Ur baseline (cross spectrum)is also presented.
 The number in the parentheses corresponds to single pulses 
	of which the pulse time range is overlapped with the pulsar phase 
	range (high probability pulsar signal). \label{tab:result_auto}}
\begin{tabular}{c|ccc|c}
	\hline\hline
	Scan No.& \multicolumn{3}{c}{Auto spectrum} & Cross spectrum \\
			& Km & Ur & Cross matching	& Km-Ur 			\\
	\hline
	69	&	6334 (151)	&	6549 (87)	&	140 (26)	&	49 (40)	\\
	71	&	6296 (129)	&	6390 (91)	&	136 (20)	&	57 (41)	\\
	73	&	6355 (141)	&	6483 (94)	&	129 (26)	&	51 (36)	\\
	\hline\hline
\end{tabular}
\end{table}

\begin{figure}
	\centering
	\includegraphics[width=\textwidth]{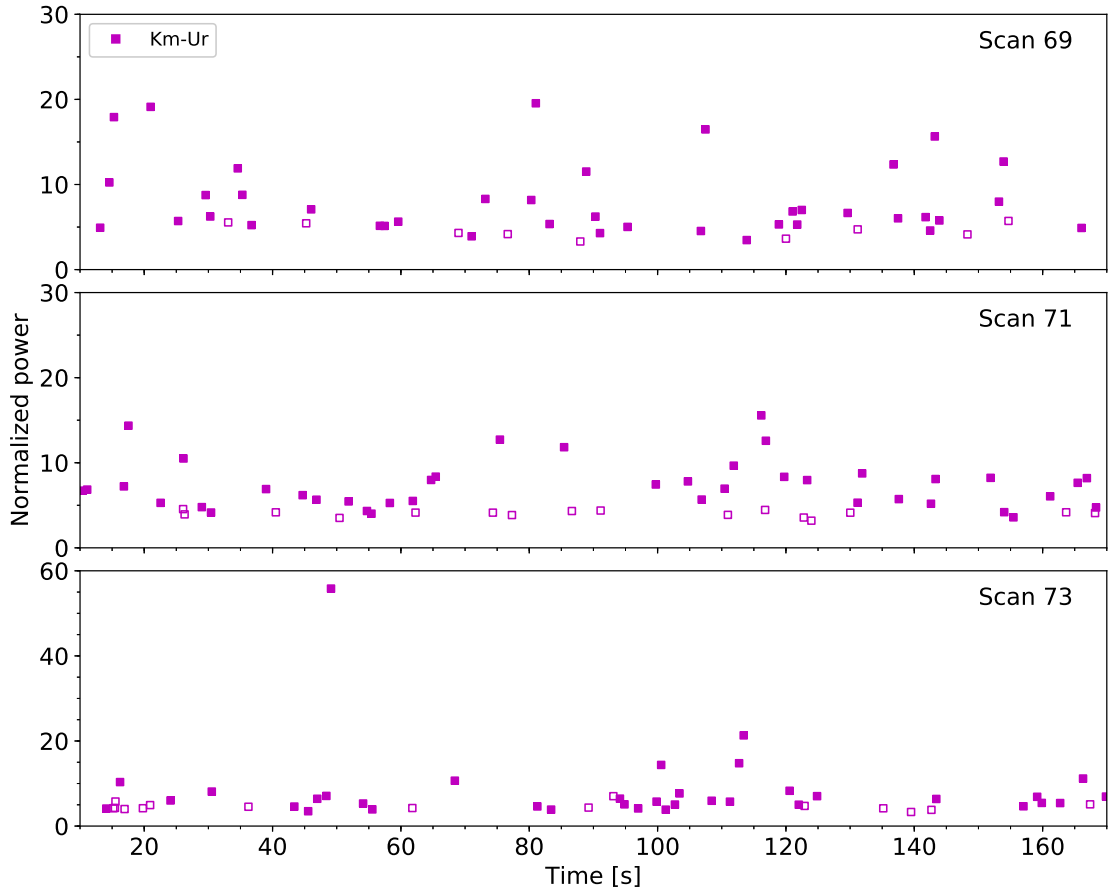}
	\caption{Cross spectrum detection result of Km-Ur baseline. Filled and 
	empty squares correspond to high possibility pulsar signals and false 
	detections, respectively. The ``normalized power'' is defined as the signal 
	power subtracted by the average level and then normalized with the standard 
	deviation (Liu et al. 2018).
		\label{fig:result_cross}}
\end{figure}

\begin{figure}
	\centering
	\includegraphics[width=\textwidth]{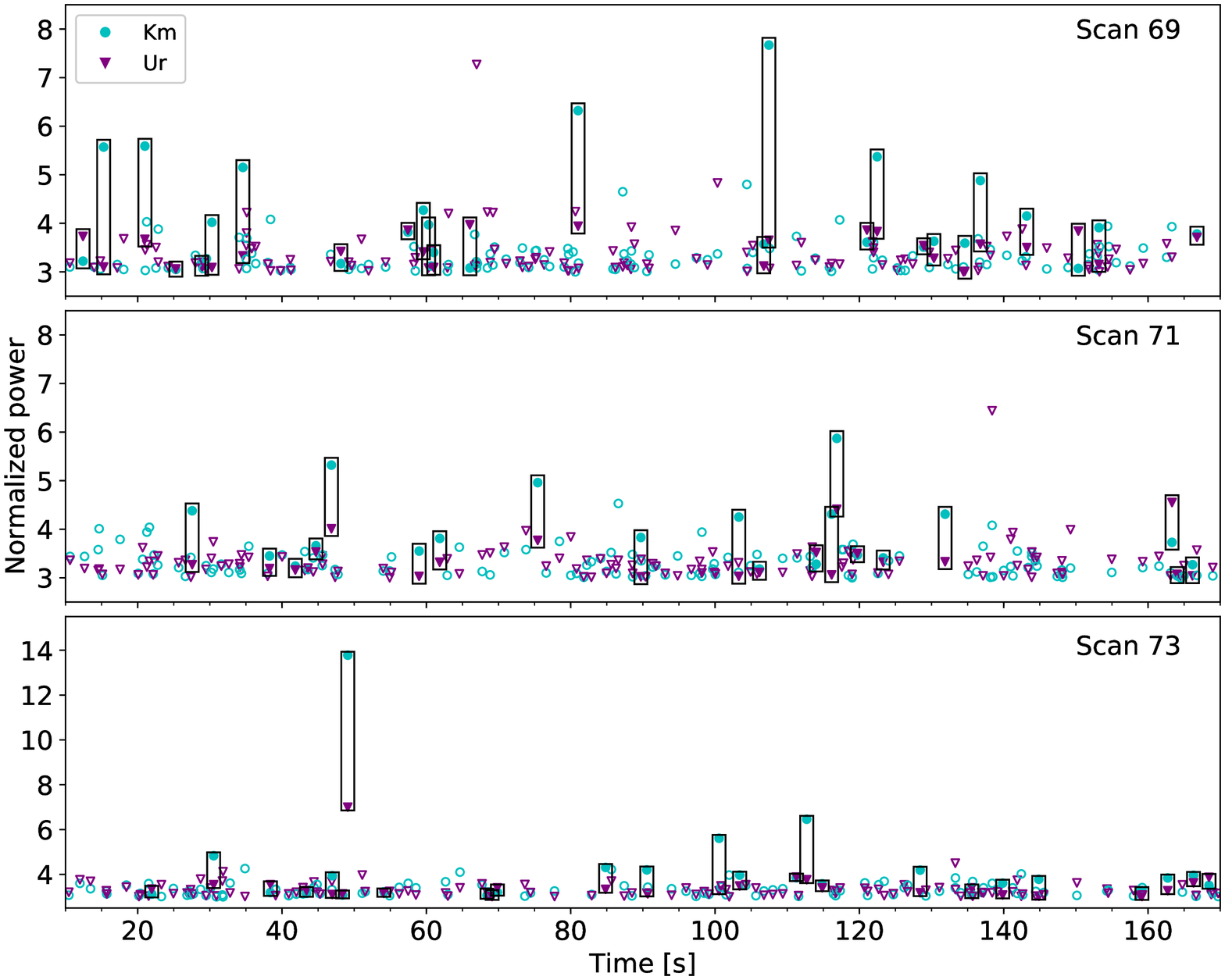}
	\caption{Cross matching result of Km and Ur stations with auto spectrum 
	method. Single pulses presented in the figure are detected 
	simultaneously (time ranges are overlapped with each other) by two 
	stations. Filled and empty symbols correspond to high possibility pulsar 
	signals and false detections, respectively.	For clarity, high possibility 
	pulsar signals are enclosed with black rectangular boxes. The ``normalized 
	power'' is defined as the detrended signal power normalized with the 
	standard deviation, as proposed in the PRESTO package.
	\label{fig:result_auto}}
\end{figure}

The Km and Ur single pulse search and two stations cross matching result
are presented in Tab. \ref{tab:result_auto}. For each station, a large number 
of single pulses are detected.
However, the detection accuracy is just slightly higher than 1\%, which 
means most of the detected signals are RFIs. By cross matching the two stations 
detection result, detection accuracy becomes higher. As a 
comparison, cross spectrum method detects more 
high possibility pulsar signals with much higher detection accuracy, which 
demonstrates that the cross spectrum based method is better 
at extracting single pulses from RFI contaminated data.

Fig. \ref{fig:result_cross} and \ref{fig:result_auto} present the Km-Ur 
baseline 
detection result with cross spectrum method and the cross matching result of
Km and Ur stations with auto spectrum method. For the auto spectrum result, the 
signal powers of Km station are usually higher than that of Ur station, which 
is consistent with their sensitivity. By comparing the two figures, we may find 
that the normalized powers of cross spectrum result are usually higher than 
that of auto spectrum result. This is because the cross spectrum method fully 
utilizes the cross spectrum fringe phase information, which enhances the signal 
power. By utilizing this feature, the cross spectrum method is able to extract 
more single pulses with higher accuracy. 

\section{Summary}\label{sec:conclusion}
In this work, we present the single pulse detection result on a VLBI pulsar 
observation data set using both cross spectrum and auto spectrum method.

Compared with auto spectrum method, cross spectrum method is able to extract 
more signal pulses with higher detection accuracy. The signal power of cross 
spectrum method is 
higher than auto spectrum method, which leads to a higher confidence level. 
The cross spectrum method is able to extract single pulses from highly RFI 
contaminated data. According to the comparison, we may find that cross spectrum 
method makes it possible to carry out FRB search in VLBI observation with low 
sensitivity telescopes and even with the presence of RFIs.

Due to the limitation of currently available data, our comparisons are only 
limited to low DM environment and do not involve DM search. It has been proved 
that auto spectrum method is very effective at excluding RFIs by large number 
of DM trials. We still have to verify the performance of the cross spectrum 
method in high DM environment. To obtain the high DM data set, a VLBI 
observation of RRAT (Rotating Radio Transit, McLaughlin et al. 2006) source is 
already in our plan. One possible choice is J1819-1458, the DM value is 196 pc 
cm$^{-3}$ and the flux is 3.6 Jy at 1.4 GHz (Keane et al. 2011).
We will present the cross spectrum method single pulse search result with this 
source in our future work.


\begin{acknowledgements}
This work was funded by the National Natural Science Foundation of China 
(11373061, 11573057, U1631122), the Key Laboratory of Radio Astronomy of 
Chinese 
Academy of Sciences, Shanghai Key Laboratory of Space Navigation and 
Positioning Techniques (ZZXT-201702), the CAS Key Technology Talent Program, 
and the 
National R\&D Infrastructure and Facility Development Program of China, 
``Fundamental Science Data Sharing Platform'' (DKA2017-12-02-09). The authors 
appreciate the support of CVN data processing center.

\end{acknowledgements}

\label{lastpage}

\end{document}